\documentclass[12pt,preprint]{aastex}

\newcommand{\kms}{km s$^{-1}$}
\newcommand{\ha}{H$\alpha$}
\newcommand{\ca}{Ca\,{\sc ii} 8542 \AA}

\slugcomment{Correspondence: M. D. Ding (dmd@nju.edu.cn)}
\shorttitle{A two-ribbon flare associated with a filament eruption}
\shortauthors{Ding et al.}

\begin{document}

\title{
H$\alpha$ and hard X-ray observations of a two-ribbon flare associated
with a filament eruption}

\author{M. D. Ding, Q. R. Chen, J. P. Li, and P. F. Chen}
\affil
  {Department of Astronomy, Nanjing University, Nanjing 210093, China}

\begin{abstract}

We perform a multi-wavelength study of a two-ribbon flare on 2002
September 29 and its associated filament eruption, observed
simultaneously in the \ha\ line by a ground-based imaging spectrograph
and in hard X-rays by RHESSI. The flare ribbons contain several \ha\
bright kernels that show different evolutional behaviors.
In particular, we find two kernels that may be the footpoints of
a loop. A single hard X-ray source appears to cover these two
kernels and to move across the magnetic neutral line. 
We explain this as a result of the merging of two footpoint
sources that show gradually asymmetric emission
owing to an asymmetric magnetic topology of the
newly reconnected loops. In one of the \ha\ kernels, we detect
a continuum enhancement at the visible wavelength. By checking
its spatial and temporal relationship with the hard X-ray
emission, we ascribe it as being caused by electron beam
precipitation. In addition, we derive the line-of-sight velocity
of the filament plasma based on the Doppler shift of the
filament-caused absorption in the \ha\ blue wing. The filament
shows rapid acceleration during the impulsive phase. These
observational features are in principal consistent with the
general scenario of the canonical two-ribbon flare model.

\end{abstract}


\section{Introduction}

Two-ribbon flares are usually accompanied by the eruption of filaments
or coronal mass ejections (CMEs). A generally accepted model for
two-ribbon flares predicts that prior to the flare, a filament
(prominence) lies above magnetic arcades, which becomes unstable owing
to some processes such as a shear motion of magnetic arcades (Miki\'c
\& Linker 1994), a converging motion of them (Forbes \& Priest 1995),
or an emerging flux (e.g., Chen \& Shibata 2000). The filament eruption
leaves below a current sheet where magnetic reconnection occurs and
energy is released to power a solar flare. Recently, the relationship
between flares and filament eruptions (or CMEs) has become a hot topic
(e.g., Zhang et al. 2001; Wang et al. 2003).

A two-ribbon flare comprises of a series of arcade loops whose footpoints
form the two flare ribbons that are most often observed in \ha.
It is known from observations with high spatial resolution that several
flare kernels can appear in one ribbon, which correspond to different
flaring loops whose energy processes (heating mechanisms) may be
different (e.g., Kitahara \& Kurokawa 1990). Asai et al. (2003)
even identified small-scale conjugate footpoints in the two ribbons
of a flare by searching for the most correlated \ha\ light curves.
Discrimination between different heating mechanisms
(electron bombardment versus heat conduction) relies on several
ways. First, as is well known, the hard X-ray emission provides a
direct information on the site where high-energy electrons precipitate.
Second, the emission features of the \ha\ line can also be used as
an indirect diagnostic tool for the heating mechanisms, based on the
line shape and width (e.g., Canfield, Gunkler, \& Ricchiazzi 1984;
Canfield \& Gayley 1987; Fang, H\'enoux, \& Gan 1993) and the comparison
of \ha\ and hard X-ray light curves (e.g., Wang et al. 2000; Trottet
et al. 2000). In a few cases, heating at deeper layers can be achieved
that produces an enhanced white-light emission. A probable cause
is chromospheric heating by a strong electron beam followed by the
backwarming effect (Liu, Ding \& Fang 2001; Ding et al. 2003).

In this paper, we perform a multi-wavelength study of a two-ribbon flare
on 2002 September 29, observed simultaneously by the Solar Tower
Telescope of Nanjing University and the Reuven Ramaty High
Energy Solar Spectroscopic Imager (RHESSI). The following topics
are in particular addressed: (1) spatial and temporal variation of
the \ha\ and continuum emission; (2) dynamics of the filament eruption
in association with the flare development; and (3) distribution of
the hard X-ray source and its variation with time.

\section{Observations}

A two-ribbon flare (importance M2.6/2N), associated with a filament
eruption, occurred at N12\,E21 in NOAA Active Region 0134, on 2002
September 29. It started from 06:32 UT and peaked at 06:39 UT. We
made spectral observations in \ha\ and \ca\ lines for this flare during
the period of 06:35--06:49 UT, using the imaging spectrograph installed
in the Solar Tower Telescope of Nanjing University (Huang et al. 1995;
Ding et al. 1999).
Two-dimensional spectra for the whole flaring region were obtained by
using a scanning technique. A frame of two-dimensional spectra
contains a three-dimensional data array: 260 wavelength points
around the line, 120 points along the slit with a pixel spacing
0$\farcs$85, and 50 points along the scanning direction
with a spacing 2$\arcsec$. The spectral resolution is 0.05 and
0.118 \AA\ pixel$^{-1}$ for \ha\ and \ca\ respectively. Therefore,
we can detect the emission at the far wings of the lines
($\Delta\lambda=6$ \AA\ for \ha\ and 15 \AA\ for \ca),
which can be served as a proxy of the continuum
window. We repeated in total 17 scans with a
repetition time about 15 s in the impulsive phase. The last scan was
done well after the flare when the continuum emission has dropped
back to the quiescent value.

This flare was also observed by RHESSI, which provides the first
high-resolution hard X-ray imaging spectroscopy (Lin et al. 2002).
The hard X-ray data yield important information on high-energy
electrons and the heating of the flare atmosphere.
In addition, we use the data from the Michelson Doppler Imager (MDI)
on board the {\it Solar and Heliospheric Observatory} ({\it SOHO})
to deduce the magnetic geometry related to this flare.

Reduction of the ground-based data includes a correction for
the dark field and the flat field. \ha\ images of the flaring region
are reconstructed from the 2D spectral data. To compare images from
different instruments, we align the \ha\ image at the line wing with
the MDI intensity image by tracing the sunspot features in the active
region. The accuracy of image alignment is estimated to be
$\sim 2\arcsec$. Other images can be overlapped similarly.

\section{Results and Discussions}

\subsection{Evolution of the flare ribbons and the emission features}

Figure 1 shows the \ha\ monochromatic images (at the line center) of
the 2002 September 29 flare at some selected times covering the
impulsive phase. Overlapped is the magnetic neutral line
from the MDI magnetogram observed at 06:24:01 UT, $\sim 15$ min prior
to the flare maximum time. Therefore, when drawing the magnetic
neutral line, we have applied a correction of 15 min solar rotation
for consistency between the \ha\ image and the magnetogram.
It is seen that the curved magnetic neutral line divides
the flare into three parts. The lower part below the neutral line is
a small (fainter) ribbon. The upper part (above the neutral line )
and the middle part (surrounded by the neutral line) look like to
form a big (brighter) ribbon. However, they lie in areas of different
magnetic polarities. In particular, we draw attention to two points
across the neutral line, labeled as ``A'' and ``B'' in Fig.~1, and plot
the time profiles of the \ha\ emission in Fig.~2. Point ``A'' is the
center of the kernel in the upper part of the flare and point ``B'' is
the center of the kernel in the middle part (also the brightest kernel
in the flare). We find that point ``A'' is already relatively hot
at the start of observations. It then cools down gradually with time.
Point ``B'', however, is relatively cool at first and is heated
rapidly during the impulsive phase.

\begin{figure}[tb]
\epsscale{1.0}
\plotone{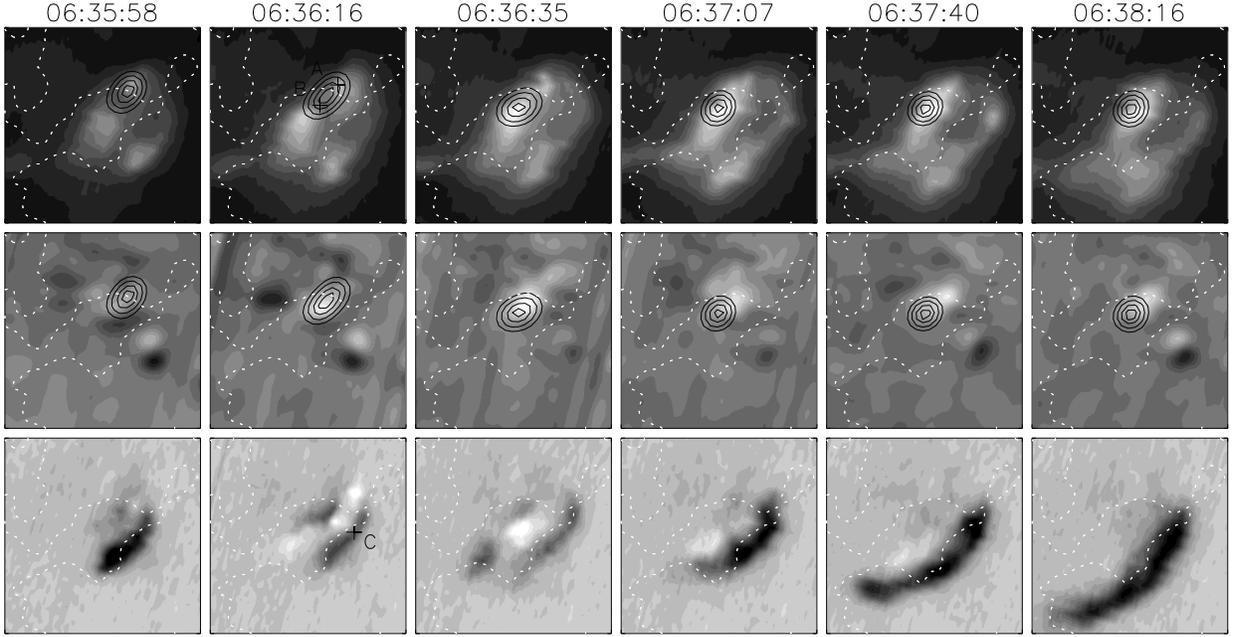}
\caption[]{Images showing the evolution of the 2002 September 29 flare.
{\it Top panel}: images at the \ha\ line center;
{\it middle panel}: images at H$\alpha +6$ \AA\ as a proxy of the
continuum emission;
and {\it bottom panel}: difference of the H$\alpha -4$ \AA\ and
H$\alpha +4$ \AA\ images showing the motion of the filament.
The RHESSI hard X-ray contours (12--25 keV) are superimposed on the
top and middle panels at each time. The magnetic neutral line,
observed by {\it SOHO}/MDI, is also plotted as a white dotted line.
``A'', ``B'', and ``C'' are selected points that we refer to in
Fig.~2. The field of view is $60\arcsec \times 60\arcsec$.
North is up, and east is to the left.}
\end{figure}

As mentioned above, a two-ribbon flare may consist of many small loops
that flare up and evolve differently, and that produce light curves
of different shapes. Generally speaking, the \ha\ line emission
originates from and reflects a heating status in the upper
chromosphere, while the white-light continuum emission is from the
lower chromosphere or photosphere. For the present event,
we also search for the continuum emission near the \ha\ line.
We select a wavelength window at the far red wing (i.e.,
6 \AA\ from the line center) to avoid a possible effect of
the filament absorption, and then derive
the relative enhancement of the intensity, defined as
$R=(I_{f}-I_{q})/I_{q}$, where $I_{f}$ is the intensity during the
flare and $I_{q}$ is the quiescent value at the same point. The
latter is taken as the intensity well after the flare ($\sim 10$
min after the flare maximum). The same is done for the continuum
near the \ca\ line, but we use the wavelength window at 15 \AA\
from the line center.
The time variations of the \ha\ line center intensity and the
continuum intensity for points ``A'' and ``B'' are plotted in
Fig.~2, together with the RHESSI hard X-ray counts summed over
detectors 1 through 9, excluding detectors 2 and 7.

We detect a fairly strong continuum enhancement ($R\approx 8$ \%)
at point ``B''. The continuum emission rises more rapidly than the
line emission; it reaches a maximum that corresponds roughly to the
peak of $\ga 25$ keV hard X-ray emission, or a sub-peak of the hard
X-ray emission at lower energies; then, it keeps above the quiescent
value until the later phase of the flare. Note that we find a similar
enhancement with a slightly less magnitude at the continuum near
the \ca\ line. The visible continuum at point ``A'' seems
also enhanced but much less significant than at point ``B''.
As will be discussed below, such a continuum emission
may be related to electron beam precipitation.

\begin{figure}[tb]
\epsscale{0.6}
\plotone{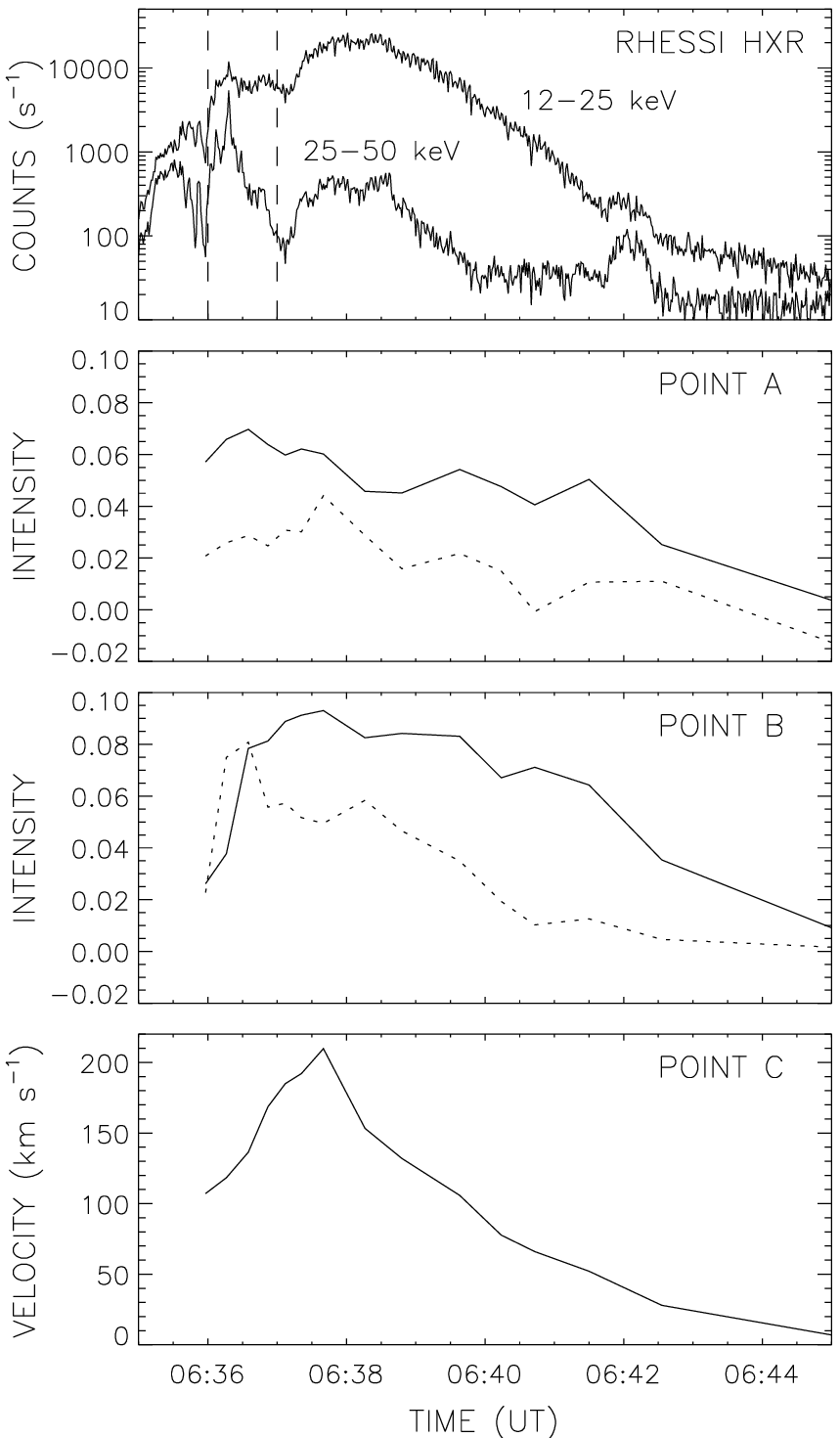}
\caption[]{Time profiles of the RHESSI hard X-ray emission at
12--25 and 25--50 keV energy bands ({\it top panel}), the net
increase of the emission at \ha\ line center ({\it solid line}) and
at H$\alpha +6$ \AA\ ({\it dotted line}), normalized to the quiescent
intensity at H$\alpha +6$ \AA, at points ``A'' and ``B''
({\it middle panels}), and the line-of-sight velocity at point
``C'' in the filament ({\it bottom panel}).
The two vertical bars in the top panel refer to the integration
time interval for which the photon spectrum is generated (Fig.~3).}
\end{figure}

We further study the origin of the hard X-ray emission by making
a fit to the RHESSI hard X-ray spectra. Figure 3 plots the spatially
integrated, background-subtracted photon spectrum for the time
interval 06:36:00--06:37:00 UT, obtained using the RHESSI spectral
executive software (Smith et al. 2002). We fit this spectrum by a
bremsstrahlung spectrum that originates from both a thermal source
and a thick-target source, as was done in Sui et al. (2002).
The thick-target source yields a power-law fit. The results show that
the non-thermal contribution to the hard X-ray emission dominates
over the thermal source when $E\ga 14$ keV; the former becomes
nearly one order of magnitude stronger than the latter at $E\sim 20$
keV.

\begin{figure}[tb]
\epsscale{1.0}
\plotone{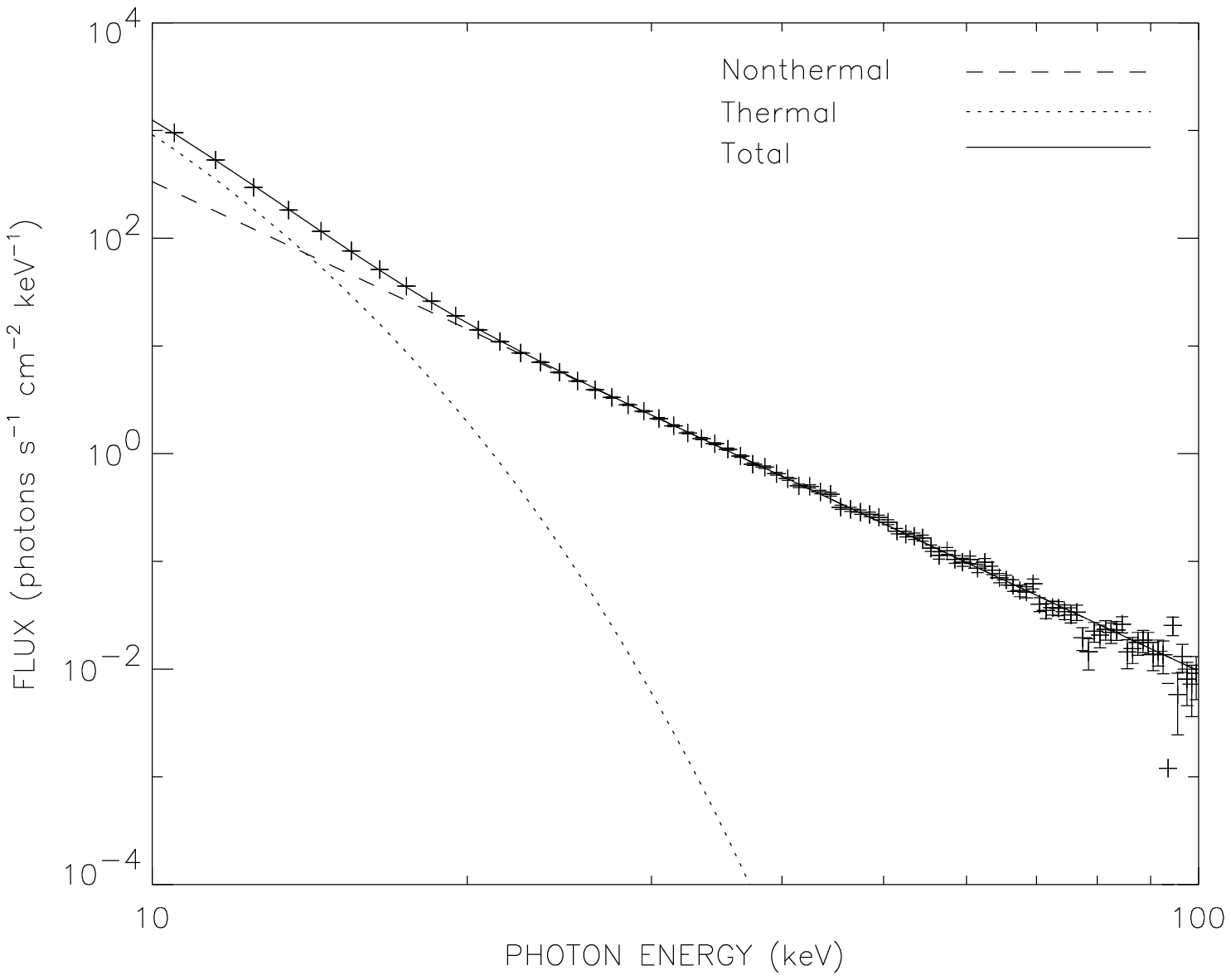}
\caption[]{RHESSI photon spectrum for the time interval
06:36:00--06:37:00 UT, fitted by the spectrum from a thermal source
plus a thick-target source.}
\end{figure}

\subsection{The filament eruption}

The 2002 September 29 flare is closely associated with a filament
eruption. In the observed \ha\ and \ca\ spectra, the filament
causes an absorption, or sometimes an emission, in the line blue wing.
To view this feature clearly, we subtract the \ha\ images at the
red wing ($\Delta\lambda=4$ \AA) from that at the blue wing
($\Delta\lambda=-4$ \AA) and plot these difference images in Fig.~1.
The figure shows clearly the location and evolution of the filament
(more accurately, the moving material in the filament).
Seen from the MDI magnetogram overlapped on the image,
the filament lies along the magnetic neutral line. With the flare
development, the filament expands along its axis, in the southeastern
direction. This phenomenon can be explained as a projection of
the motion of the filament mass in the plane perpendicular to the
line of sight. Since we cannot determine the velocity vector,
we are unable to trace a specific mass element and study its
dynamics. Therefore, we try to deduce the line-of-sight velocity
for a fixed spatial point within the filament, based on the Doppler
shift of the absorption feature in the line wing. The time history
of the line-of-sight velocity for a typical point in the midst of
the filament, marked as ``C'' in Fig.~1, is shown in Fig.~2.
The filament is shown to undergo a drastic
acceleration during the impulsive phase; it reaches a maximum
line-of-sight velocity of $\sim 210$ \kms\ at $\sim$06:37:40 UT,
roughly in coincidence with the \ha\ maximum at point B;
after this, the filament is decelerated, while it
continues to rise higher until the later phase of the flare.

The relationship between filament eruptions and flares has been
extensively studied for decades (e.g., Rust 1976;
Hanaoka et al. 1994; Wang et al. 2003). For a limb event,
one can accurately determine
the height of the erupting filament (prominence) and derive its
propagation speed (e.g., Klein \& Mouradian 2002). Since the event
studied here is close to the disk center, it is hard to measure
the height of the filament. The velocity derived from the
Doppler shift of the absorption center reflects only a mean
velocity averaged over the line of sight; it more probably
corresponds to the velocity in the base part of the filament
where the mass density is higher and most absorption is produced.
However, the fact that the acceleration phase of the filament
eruption corresponds to the flare impulsive phase provides a
convincing confirmation of the result of Zhang et al. (2001).

We also notice from Fig.~1 ({\it the lower panels}) that the filament
does not always appear in absorption (dark features); in stead, 
some parts may appear in emission (bright features) in the \ha\
blue wing. This implies either an ejection of heated
plasma (Batchelor \& Hindsley 1991) or a heating of the filament
material during its eruption. Note that such an emission feature
appears especially in the impulsive phase.

\subsection{The hard X-ray source and its motion}

The spatial distribution and the temporal evolution of the hard X-ray
emission provide clues to the acceleration of energetic electrons
and to the heating of the flaring atmosphere. When a flare occurs,
electrons are accelerated in the magnetic reconnection site and then
precipitate downward along the magnetic field lines to both footpoints
of the flaring loop, producing hard X-ray emission there. Therefore, in
most cases, flares exhibit double hard X-ray sources (Sakao 1994), or
multiple sources in the case of more than one flaring loops. However,
in quite a few cases, only a single source in seen. This is either
because the two footpoint sources are too close to be spatially
resolved, or due to an asymmetric magnetic topology, in which the
magnetic field in one loop leg is so converging that the magnetic
mirroring effect inhibits most electrons from streaming downward.

\begin{figure}[tb]
\epsscale{1.0}
\plotone{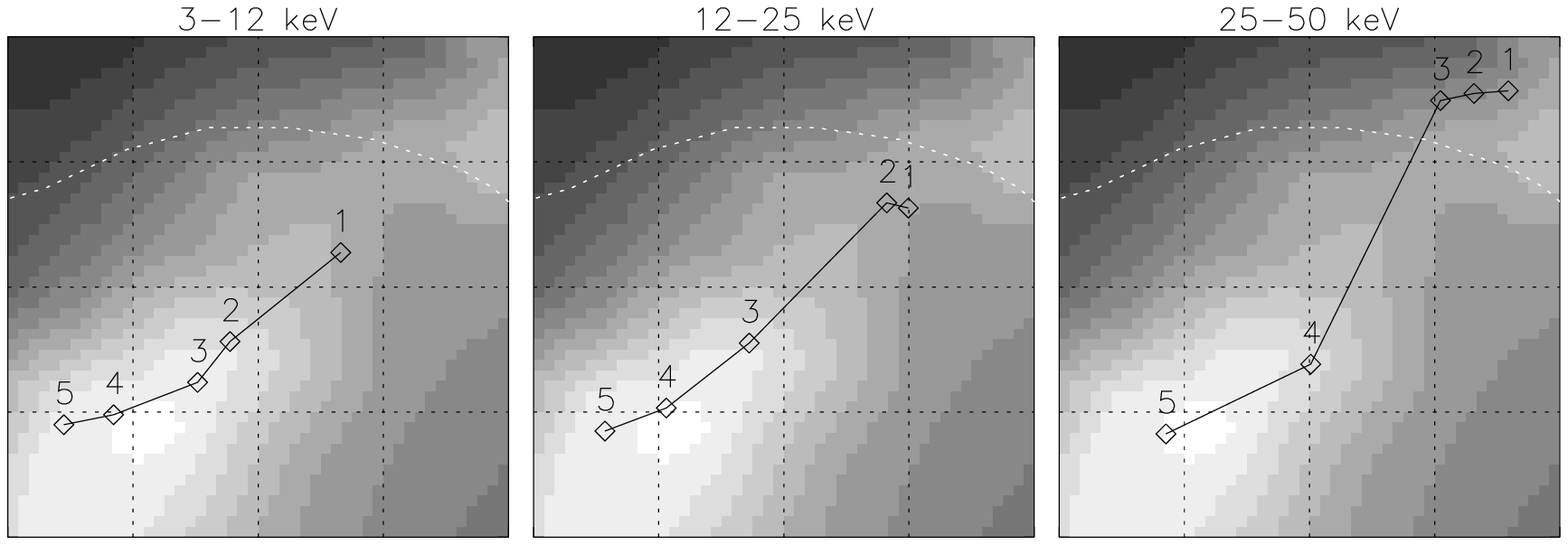}
\caption[]{Position of the centroid of the main hard X-ray source
at 5 consecutive time intervals of 12 s between 06:36:00 and
06:37:00 UT ({\it diamonds}), and the magnetic neutral line
({\it white dotted line}) overlaid on the flare image
at 06:36:35 UT. The field of view is $12\arcsec \times 12\arcsec$.}
\end{figure}

The RHESSI data provide a good opportunity to study the above issue. To
this end, we construct the hard X-ray images at different times that are
overlaid on the \ha\ images in Fig.~1.
In producing the hard X-ray images, we adopt the clean procedure,
set the integration time to 12 s, and include detectors 3 through 8.
We notice that a strong hard
X-ray source appears, covering the bright points ``A'' and ``B'', from
the impulsive phase to the gradual phase. In the fainter ribbon,
however, we only detect a rather weak source at the very beginning
($\sim$ 06:35--06:36 UT), prior to the \ha\ observation time. This weak
source is of a very short duration, located in the southeastern part of
the fainter ribbon, which seems to have no relation to
the strong source. Therefore, we think that the strong source itself
contains two footpoint sources that are spatially mixed. A fact
supporting this point is that the source straddles over the magnetic
neutral line at earlier times (Fig.~1).

A further inspection of Fig.~1 reveals a motion of the hard X-ray
source across the magnetic neutral line in the early phase.
For a quantitative view, we plot in Fig.~4 the position of the
centroid of the source at different energy bands for 5 successive
time intervals of 12 s during 06:36--06:37 UT.
Note that the centroid is defined as the intensity-weighted center
of the hard X-ray source with $\ge 50$\% peak intensity. Since the
shape of the hard X-ray source is nearly round, the centroid is
roughly the place of peak intensity.
The source motion is seen to depend on the energy of the emission.
The source at a higher energy band moves later, but across a
longer distance, than that at a lower energy band.
Note that such a difference can be explained by the different
proportions of thermal and non-thermal sources at different energy
bands. As revealed in Fig.~3, the hard X-ray emission at 3--12 keV
is mainly thermal, while that at 25--50 keV is nearly nonthermal;
at 12--25 keV, the thermal and nonthermal contributions are
comparable.
(In fact, we have also tried to construct the hard X-ray image
at the 50--100 keV range; however, the centroid of the source
cannot be reliably determined since the count rates are not
large enough.) To trace the fastest source motion,
we shorten the time interval (also the integration time) to as
small as 2 s; at the 25--50 keV band, the biggest motion during
this small interval is $\ga 2\arcsec$, which means a source motion
speed as large as $\ga 1\arcsec$
per second. We also find that at first, the sources at different
energy bands are more spatially separated than later;
they converge to nearly the same place and become relatively
stable after their motions stop (at $\sim$ 06:37 UT).

Sakao et al. (2000) have studied in detail the motion of hard X-ray
(double) sources in flares observed by {\it Yohkoh}. They found that
in most cases, the double sources move anti-parallelly.
Krucker, Hurford, \& Lin (2003) also found a hard X-ray source
motion roughly parallel to the magnetic neutral line with a
velocity of up to $\sim 100$ \kms. Masuda, Kosugi, \& Hudson (2001)
found both kinds of hard X-ray source motions, parallel and
vertical to the magnetic neutral line, for an X-class flare.
In our case, the source motion does show both the parallel and
vertical components.
Note that at the 25--50 keV energy band, the motion is nearly
perpendicular to the magnetic neutral line during the period of
fast motion. Since this source is probably a merging of two
footpoint sources, we propose two causes for such a motion:
an asymmetric hard X-ray emission, that is, the footpoint source
at the southern side of the neutral line gradually dominates
over that at the northern side, and/or a ribbon separation from
the neutral line (e.g., Qiu et al. 2002). We think that the first
effect is more important during the short period of fast motion.
That is to say, there is a mechanism that suddenly switches off
(or greatly reduces) the emission from the northern side but enhances
the emission from the southern side. We further postulate that the
magnetic topology of the newly reconnected loops becomes very
asymmetric so that the magnetic mirroring effect results in such an
asymmetric hard X-ray emission.

To confirm the above point, we compute the magnetic topology
of the active region from the MDI magnetogram using a potential
field extrapolation developed by Sakurai (1982). In Fig.~5, we
schematically plot the magnetic field lines projected on the MDI
magnetogram. For clarity, only the field lines in the flaring
region are given. The connectivity of the magnetic field indicates
that there are magnetic loops that straddle over the upper part of
the magnetic neutral line. These loops are responsible for the
production of the main hard X-ray source and the continuum
emission, as discussed above. In addition, there appear other
loops that connect the \ha\ brighter ribbon and the fainter ribbon
(across the lower part of the magnetic neutral line).
The overall picture of this flare is a sequence of sheared magnetic
arcade loops and a filament lying above them. Unfortunately, the
cadence of MDI magnetograms is only 96 min; therefore, we are
unable to learn the magnetic field change during the flare.

Note that we do not find a corresponding motion of the continuum
emission source. This is not surprising if we interpret the
continuum emission as due to the backwarming effect (Ding et al. 2003).
The lower atmosphere requires tens of seconds to get fully heated,
so that the continuum emission cannot follow closely in space the
fast motion of the hard X-ray emission.
Of course, it may also be due to the relatively low spatial
resolution of the 2D spectral observations.

\begin{figure}[tb]
\epsscale{1.0}
\plotone{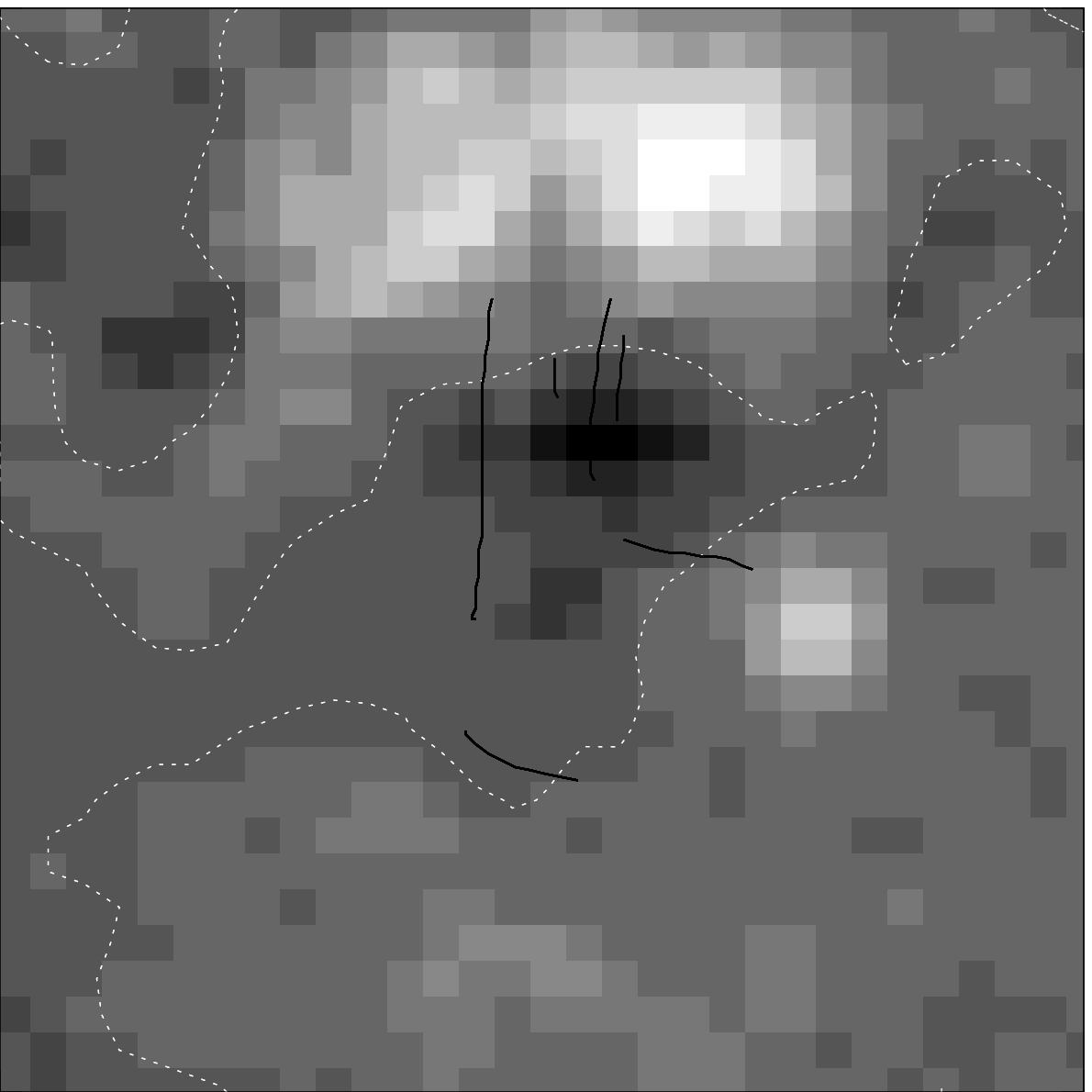}
\caption[]{Magnetic field lines computed using a potential field
extrapolation from the MDI data ({\it solid line}). Also plotted
are the MDI magnetogram ({\it gray scale}) and the magnetic neutral
line ({\it white dotted line}). The field of view is
$60\arcsec \times 60\arcsec$.}
\end{figure}

\section{Conclusions}

We have studied the \ha\ emission and the RHESSI hard X-ray emission
of the 2002 September 29 flare. The main results are as follows.

(1) The flare is associated with a filament eruption. A rapid
acceleration of the filament occurs in the flare impulsive phase.
The line-of-sight velocity deduced at a fixed
spatial point within the filament is seen to reach a maximum
at nearly the same time of \ha\ maximum in the brightest kernel.
In the impulsive phase, some parts of the filament may produce
an emission in the \ha\ blue wing. This implies either a heating
of the filament mass or an ejection of heated plasma.

(2) The flare ribbons comprise of several \ha\ kernels that evolve
differently. Judging from the magnetic topology and the emission
features in both \ha\ and hard X-rays, we have identified two bright
points, which lie at both sides of the magnetic neutral line and are
probably the footpoints of a flaring loop.

(3) We detect a motion of the hard X-ray source across the
magnetic neutral line. Since this source is probably a merging of
two footpoint sources, its motion is ascribed to a rapid change of
their relative weights owing to a change of the magnetic topology
of the newly reconnected loops. However, it is still unclear what is
the mechanism that produces such a change in a very short time
($\sim 2$ s).

(4) We also detect an enhanced emission at the visible
continuum (near \ha) and the infrared continuum (near \ca), at a
place near the center of the hard X-ray source. Liu et al.
(2001) have reported a similar continuum emission in a flare
of 2001 March 10. Ding et al. (2003) proposed a model for the
continuum emission as a result of electron beam heating followed by
radiative backwarming. The results here favor such an interpretation.

In principal, the observed features of this flare can still be
interpreted by the canonical model of two-ribbon flares
(Kopp \& Pneuman 1976).

\acknowledgments

We would like to thank the referee for his/her valuable comments
on the paper. We are grateful to the RHESSI
team and the {\it SOHO}/MDI team for providing the observational data,
and to T. Sakurai for providing the computer code for magnetic field
extrapolation. This work was supported by NKBRSF under grant G20000784,
by NSFC under grants 10025315 and 10221001, and by a grant from TRAPOYT.


\begin{thebibliography}{}

\bibitem{} Asai, A., Ishii, T. T., Kurokawa, H., Yokoyama, T.,
  \& Shimojo, M. 2003, ApJ, 586, 624
\bibitem{} Batchelor, D. A., \& Hindsley, K. P. 1991, Sol. Phys., 135, 99
\bibitem{} Canfield, R. C., \& Gayley, K. G. 1987, ApJ, 322, 999
\bibitem{} Canfield, R. C., Gunkler, T. A., \& Ricchiazzi, P. J. 1984,
  ApJ, 282, 296
\bibitem{} Chen, P. F., \& Shibata, K. 2000, ApJ, 545, 524
\bibitem{} Ding, M. D., Fang, C., Yin, S. Y., \& Chen, P. F. 1999,
  A\&A, 348, L29
\bibitem{} Ding, M. D., Liu, Y., Yeh, C.-T., \& Li, J. P. 2003, A\&A,
  403, 1151
\bibitem{} Fang, C., H\'enoux, J.-C., \& Gan, W. Q. 1993, A\&A, 274, 917
\bibitem{} Forbes, T. G., \& Priest, E. R. 1995, ApJ, 446, 377
\bibitem{} Hanaoka, Y., et al. 1994, PASJ, 46, 205
\bibitem{} Huang, Y. R., Fang, C., Ding, M. D., Gao, X. F., Zhu, Z. G.,
  Ying, S. Y., Hu, J., \& Xue, Y. Z. 1995, Sol. Phys., 159, 127
\bibitem{} Kitahara, T., \& Kurokawa, H. 1990, Sol. Phys., 125, 321
\bibitem{} Klein, K.-L., \& Mouradian, Z. 2002, A\&A, 381, 683
\bibitem{} Kopp, R. A., \& Pneuman, G. W. 1976, Sol. Phys., 50, 85
\bibitem{} Krucker, S., Hurford, G. J., \& Lin, R. P. 2003, ApJ, in press
\bibitem{} Lin, R. P., et al. 2002, Sol. Phys., 210, 3
\bibitem{} Liu, Y., Ding, M. D., \& Fang, C. 2001, ApJ, 563, L169
\bibitem{} Masuda, S., Kosugi, T., \& Hudson, H. S. 2001, Sol. Phys.,
  204, 55
\bibitem{} Miki\'c, Z., \& Linker, J. A. 1994, ApJ, 430, 898
\bibitem{} Qiu, J., Lee, J., Gary, D. E., \& Wang, H. 2002, ApJ, 565, 1335
\bibitem{} Rust, D. M. 1976, Sol. Phys., 47, 21
\bibitem{} Sakao, T. 1994, Ph.D. thesis, Univ. Tokyo
\bibitem{} Sakao, T., Kosugi, T., Masuda, S., \& Sato, J. 2000,
  Adv. Space Res., 26, 497
\bibitem{} Sakurai, T. 1982, Sol. Phys., 76, 301
\bibitem{} Smith, D. M., et al. 2002, Sol. Phys., 210, 33
\bibitem{} Sui, L., Holman, G. D., Dennis, B. R., Krucker, S.,
  Schwartz, R. A., \& Tolbert, K. 2002, Sol. Phys., 210, 245
\bibitem{} Trottet, G., Rolli, E., Magun, A., Barat, C., Kuznetsov, A.,
  Sunyaev, R., \& Terekhov, O. 2000, A\&A, 356, 1067
\bibitem{} Wang, H., Qiu, J., Denker, C., Spirock, T., Chen, H.,
  \& Goode, P. R. 2000, ApJ, 542, 1080
\bibitem{} Wang, H., Qiu, J., Jing, J., \& Zhang, H. 2003, ApJ, in press
\bibitem{} Zhang, J., Dere, K. P., Howard, R. A., Kundu, M. R.,
  \& White, S. M. 2001, ApJ, 559, 452

\end{thebibliography}
\end{document}